%% file: main.tex
\documentclass[10pt,conference]{IEEEtran}
\IEEEoverridecommandlockouts
\usepackage{cite}
\usepackage{amsmath,amssymb,amsfonts}
\usepackage{algorithmic}
\usepackage{graphicx}
\usepackage{textcomp}
\usepackage{comment}
\usepackage{xcolor}
\def\BibTeX{{\rm B\kern-.05em{\sc i\kern-.025em b}\kern-.08em
    T\kern-.1667em\lower.7ex\hbox{E}\kern-.125emX}}
\usepackage{array}
\usepackage{tabularx}
\usepackage{booktabs}   %
\usepackage{listings}

\newcolumntype{Y}{>{\raggedright\arraybackslash}X}
\newcolumntype{Z}{>{\centering\arraybackslash}X}

\usepackage{xcolor}
\usepackage{listings}

\newcommand{\proplabel}[1]{%
    \textcolor{magenta!70!black}{\bfseries(#1)}%
}

\lstdefinestyle{bmcprops}{
    language=C,
    basicstyle=\ttfamily\scriptsize,
    keywordstyle=\color{blue!70!black}\bfseries,
    commentstyle=\color{teal!65!black}\itshape,
    stringstyle=\color{orange!65!black},
    morekeywords={ensures,assert,assert_eq,__CPROVER_assert,should_panic},
    backgroundcolor=\color{gray!3},
    frame=single,
    rulecolor=\color{gray!35},
    columns=fullflexible,
    keepspaces=true,
    breaklines=true,
    showstringspaces=false,
    escapeinside={(*@}{@*)},
    xleftmargin=5pt,
    xrightmargin=5pt,
    framexleftmargin=2pt,
    framexrightmargin=2pt,
    captionpos=t,
    aboveskip=6pt,
    belowskip=5pt
}

\newcommand{\data}[1][]{\textcolor{red}{\textbf{[data]}}}

\graphicspath{{figures/}{./}}
\newif\ifDEBUG
\DEBUGfalse

\newif\ifARXIV
\ARXIVfalse
\ARXIVtrue

\newif\ifCUSTOMWATERMARK
\CUSTOMWATERMARKtrue
\CUSTOMWATERMARKfalse

\input{typesetting}
\usepackage{cleveref}

\begin{document}

\title{Small Harnesses, Default Checks, Limited Coverage: Bounded Model Checking in Practice}

\ifARXIV
\author{\IEEEauthorblockN{Kevin Kolyakov}
\IEEEauthorblockA{
\textit{University of Waterloo}\\
Waterloo, Canada \\
kkolyakov@uwaterloo.ca}
\and
\IEEEauthorblockN{Ricardo Calvo}
\IEEEauthorblockA{\textit{Purdue University}\\
West Lafayette, USA \\
rcalvome@purdue.edu}
\and
\IEEEauthorblockN{James C. Davis}
\IEEEauthorblockA{\textit{Purdue University}\\
West Lafayette, USA \\
davisjam@purdue.edu}
\and
\IEEEauthorblockN{Paschal C. Amusuo}
\IEEEauthorblockA{\textit{Purdue University}\\
West Lafayette, USA \\
pamusuo@purdue.edu}
}
\else
\author{\IEEEauthorblockN{Anonymous Authors}}
\fi

\maketitle

\begin{abstract}
The growing use of AI to generate software and verification artifacts has increased the need for automated systems that provide reliable software assurance.
Bounded model checking (BMC) offers a practical foundation for such systems by automatically checking correctness and security properties directly on software.
However, effective verification requires choices about properties, input assumptions, function models, and execution bounds that influence the guarantees obtained.
Little systematic evidence exists on how practitioners make these choices, limiting the empirical foundation for developing automated verification systems.

We present an empirical study of BMC practices in open-source C and Rust projects using CBMC and Kani.
Through LLM-assisted open coding, static analysis, and artifact execution, we characterize verification properties, artifact complexity, and verification outcomes.
Our findings reveal diverse verification objectives and generally small harnesses, but also property-specification errors and extensive use of execution bounds and environment models.
Most artifacts complete verification quickly and achieve high coverage within their scope, while repository-level coverage decreases with project size.
Our findings and annotated dataset provide empirical baselines for verification automation and motivate improved techniques for property validation, harness generation, and scaling verification breadth.
\end{abstract}


\section{Introduction}

Rapid AI-assisted code generation and advances in AI-assisted formal verification are increasing demand for automated systems that establish software correctness and security~\cite{fuSecurityWeaknessesCopilotGenerated2025,pirzada_llm-generated_2024}.
Such systems must keep pace with software production while uncovering subtle defects that threaten reliability and security.
However, their effectiveness depends on choices about which properties to verify, how to model software behavior and its environment, and how to configure verification tools.
Understanding how practitioners make these choices is therefore important for designing automated verification systems that provide meaningful software assurance.

Bounded model checking (BMC)~\cite{clarke_bounded_2001} is particularly suited to such automation because it verifies software properties with limited manual proof guidance.
Unlike techniques requiring separately constructed system models or interactive proofs, BMC tools operate directly on existing software and automatically check built-in safety properties alongside custom specifications.
Prior studies have demonstrated BMC's effectiveness in uncovering high-impact vulnerabilities~\cite{amusuo_unit_2025-1} and its practicality in industrial software development~\cite{chong_code-level_2020,wu_verifying_2024}.
However, little empirical evidence exists on the properties and other verification choices practitioners make, leaving researchers without a clear empirical basis for automating these decisions.

This paper addresses this gap through an empirical study of BMC practices in open-source software projects.
We analyze 2,005 verification artifacts from 24 C and Rust projects using two popular BMC tools, CBMC~\cite{kroening_cbmc_2014} and Kani~\cite{kani_getting_nodate}.
We employ LLM-assisted open coding to characterize the properties practitioners specify and static analysis to measure the complexity and configurations of their verification artifacts.
We also execute these artifacts to evaluate their verification results, execution costs, and code coverage.

Our study produces several findings that inform the design and evaluation of automated verification systems.
We identify 12 categories of custom verification properties and specification errors in 35 harnesses across ten projects, highlighting the need for property validation.
Although harnesses are generally small (median seven lines), 63\% specify input preconditions, 54\% configure loop bounds, and 31.2\% use function models, all of which influence verification scope.
Most artifacts (97.2\%) complete within resource limits with high artifact-level coverage (median 98\%), but project-level coverage decreases with project size, providing practical baselines for automation while exposing challenges in verification breadth.

In summary, we make three contributions:
\begin{itemize}
    \item A taxonomy of verification properties and empirical measurements of their adoption across BMC projects.
    \item An empirical characterization of the complexity, execution costs, and verification outcomes of BMC artifacts.
    \item A curated dataset of 2,005 BMC artifacts from 24 projects, providing a foundation for future empirical studies and evaluation of automated verification techniques.
\end{itemize}

\myparagraph{Significance}
Our study advances the development of automated verification systems by providing empirical evidence of existing BMC practices and their limitations.
The findings and dataset provide practical guidance and evaluation baselines for property inference, verification target selection, and automated harness generation.
They also motivate techniques for validating generated properties and assumptions, ensuring that successful verification establishes meaningful guarantees.
These advances are increasingly important as software engineers and AI agents produce code at greater speeds, creating demand for scalable and trustworthy software assurance.

\section{Background and Related Work}

This section introduces the theoretical and practical foundations for our work.

\subsection{Bounded Model Checking (BMC)}

\subsubsection{As a Verification Technique}

Bounded model checking (BMC)~\cite{clarke_bounded_2001} is an automated verification technique that checks software properties within bounded program executions.
It encodes program behaviors and the negation of target properties into a propositional formula and uses constraint solving to check whether the target properties can be satisfied.
\begin{equation}
    \Phi_{b,M}(\tau) = E_{b,M}(\tau) \land \neg P(\tau),
    \label{eq:bmc}
\end{equation}
Given a program whose executions are encoded by $E_{b,M}(\tau)$ under execution bounds $b$ and environment model $M$, BMC checks whether any execution sequence $\tau$ violates the target property $P(\tau)$ by testing the satisfiability of $\Phi_{b,M}(\tau)$.
A satisfying assignment produces a counterexample indicating a 
possible property violation, while unsatisfiability establishes that the property holds for all modeled executions.

To control verification complexity, practitioners can bound loop iterations and recursion depth ($b$) or replace callers or dependent functions with simpler behavioral models ($M$).
However, insufficient bounds and overly restrictive models can exclude executions containing real defects, while overly permissive models may introduce spurious violations~\cite{takhar_memory-safety_2025,amusuo_unit_2025-1}.
Thus, verification guarantees depend on the properties, bounds, and environment models selected.

\subsubsection{Tools}
\label{subsec:bg-bmc-tools}

Several BMC tools support mainstream programming languages, including CBMC~\cite{kroening_cbmc_2014} and ESBMC~\cite{gadelha_esbmc_2018} for C, JBMC~\cite{cordeiro_jbmc_2019} for Java, and Kani~\cite{delmasKaniModelChecker2026} for Rust.
These tools translate source programs into verification conditions for satisfiability solvers and automatically check common safety properties, including memory safety, arithmetic overflow, and assertion violations.
Practitioners can also specify custom properties through assertions and function contracts and configure which built-in checks are enabled.

Compared to other verification tools, BMC offers a tradeoff between verification effort and the scope of its guarantees.
Unlike deductive verifiers~\cite{filliatre_deductive_2011, leinoDafnyAutomaticProgram2010} and interactive theorem provers~\cite{wenzel_isabelle_2008, brendelSynthesizingImplicationLemmas2025, aitkenInteractiveTheoremProving1998}, BMC does not require separately-constructed formal system models, makes function and loop invariants optional, and does not require manual proof guidance.
However, BMC only supports properties in propositional and first order logic and its guarantees is limited by execution bounds and environment assumptions.

\subsection{Component-Level BMC and Harnesses}
\label{subsec:bg-bmc-harnesses}

To manage verification complexity, practitioners often verify individual functions or groups of functions rather than entire systems.
This component-level approach resembles unit testing and has been termed unit proofing~\cite{amusuo_unit_2025-1}.
Practitioners construct verification harnesses that initialize symbolic inputs, constrain their values through assumptions, invoke target functions, and assert expected properties.
\Cref{lst:bg-bmc-artifacts} illustrates these steps using CBMC and Kani harnesses for an increment function.

\input{listing/bmc-artifacts}

In the self-contained harnesses in \cref{lst:bg-bmc-artifacts}, the CBMC and Kani examples create symbolic inputs (line 11), restrict them to values below 100 (line 12), and invoke \texttt{inc()} before asserting that its result equals the input plus one.
These assumptions define the verification domain, while the assertions specify the properties being verified.
The listing also illustrates function contracts (CBMC lines 2--7; Kani lines 2--5), which express the same requirements as preconditions and postconditions.
The contract-checking harnesses invoke \texttt{inc()} to verify these requirements using CBMC's contract-enforcement option or Kani's \texttt{proof\_for\_contract} annotation.

Beyond input models and properties, harnesses may replace dependent functions with custom models and configure global or loop-specific unwinding bounds.
Verified function contracts can also abstract function implementations when verifying their callers, while loop contracts support invariant-based reasoning without relying solely on finite unwinding~\cite{delmasKaniModelChecker2026}.
Together with build scripts and tool configurations, these constructs form the verification artifacts that determine the scope, guarantees, and cost of each verification task.

\begin{figure*}
\centering
\includegraphics[width=0.8\linewidth]{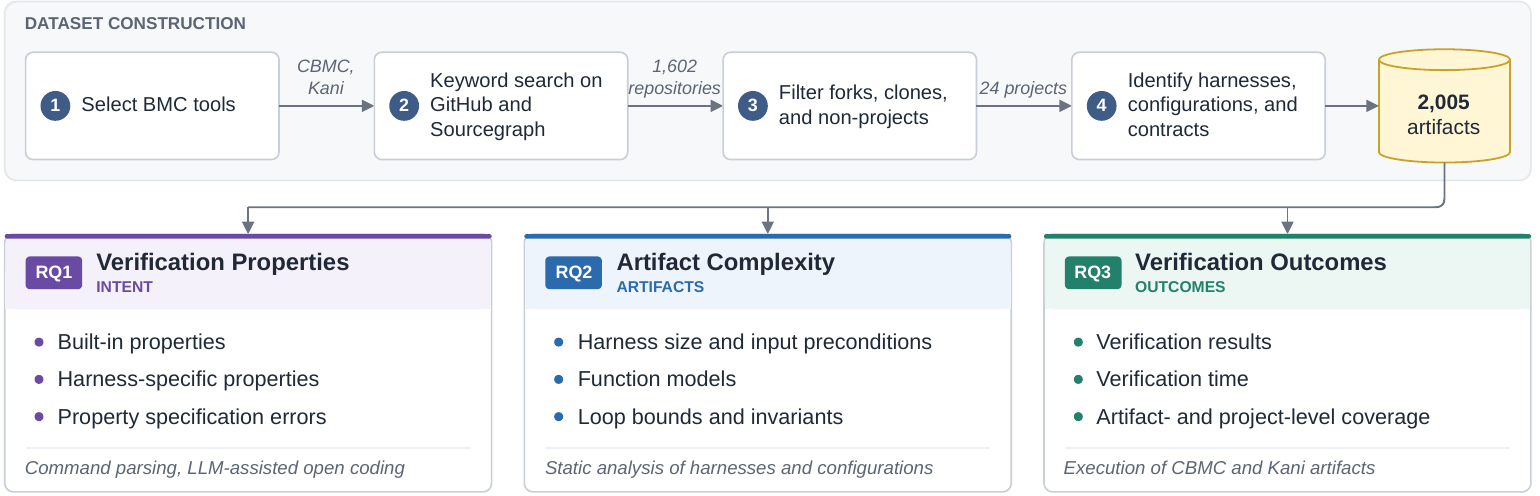}
\caption{Overview of our methodology. We select projects that contain CBMC and Kani artifacts, identify harnesses within these projects, and study their verification properties, artifact complexity and verification outcomes.}
\label{fig:methodology}
\end{figure*}

\subsection{Related Work on BMC}

We distinguish prior work improving and evaluating BMC from studies of verification practices in software projects.

\subsubsection{BMC Techniques and Applications}
Prior research has focused on improving BMC techniques, evaluating their effectiveness, and documenting their applications in specific settings.
For example, industrial studies have demonstrated BMC's practicality in software development~\cite{chong_codelevel_2021, wu_verifying_2024}, while other studies have evaluated its effectiveness in vulnerability discovery~\cite{amusuo_unit_2025-1} and improvements to verification performance~\cite{schrammel_incremental_2017}.
Recent research has also explored automating verification-task extraction~\cite{beyerHarnessForgeAutomatedExtraction2026} and specification generation~\cite{wang_kapilot_2026}.
Although these studies demonstrate BMC's capabilities and opportunities for improvement, they provide limited evidence of BMC use across independent software projects.

\subsubsection{Empirical Studies of Verification Practices}
Another line of research empirically investigates verification artifacts and practices in existing software projects.
For example, Matichuk~\etal~\cite{matichuk_empirical_2015} studied relationships between specification and proof sizes in Isabelle, while Estler~\etal~\cite{estlerContractsPractice2014a} and Dietrich~\etal~\cite{dietrichContractsWildStudy2017} examined software contract usage across projects.
Other studies have investigated program characteristics associated with verification-tool performance~\cite{demyanova_empirical_2017}.
However, these studies do not cover the verification choices and artifacts relevant to BMC across software projects.
Our study bridges these lines of research by providing empirical evidence of existing BMC practices to guide the development and evaluation of automated verification systems.

\section{Knowledge Gap and Research Questions}

Developing effective automated BMC systems requires making verification choices that establish meaningful software correctness and security guarantees.
Studying the choices practitioners make can reveal the capabilities and limitations of existing practices and provide an empirical basis for improving verification automation.
Yet, no prior study has systematically examined these practices across diverse software projects.

We address this gap through three research questions:

\begin{itemize}
    \item \textbf{RQ1 (Properties):} What do open-source projects verify?
    \item \textbf{RQ2 (Artifact Complexity):} What are the characteristics and complexity of verification artifacts used in practice?
    \item \textbf{RQ3 (Verification Outcomes):} What verification results, execution costs, and code coverage do artifacts achieve?
\end{itemize}

Answering these questions can guide automated BMC systems that select meaningful properties, choose effective verification options, and produce understandable guarantees.

\begin{table*}[t]
    \centering
    \caption{Projects in this study, with their domain, size, GitHub stars (fetched 2026-10-27), and artifact count. Size counts non-blank, non-comment lines of code (cloc) at the studied commit, excluding data/markup files and vendored dependencies.}
    \label{tab:candidate-projects}
    \footnotesize
    \setlength{\tabcolsep}{4pt}
    \begin{tabularx}{\textwidth}{@{}>{\hsize=0.85\hsize\raggedright\arraybackslash\hyphenpenalty=10000}X>{\hsize=1.2\hsize\raggedright\arraybackslash\hyphenpenalty=10000}Xrrr@{\hspace{1.5em}}>{\hsize=0.8\hsize\raggedright\arraybackslash\hyphenpenalty=10000}X>{\hsize=1.15\hsize\raggedright\arraybackslash\hyphenpenalty=10000}Xrrr@{}}
    \toprule
    \multicolumn{5}{c}{\textbf{CBMC (12 projects, 1,320 artifacts)}} & \multicolumn{5}{c}{\textbf{Kani (12 projects, 685 artifacts)}} \\
    \cmidrule(r{1.5em}){1-5} \cmidrule{6-10}
    \textbf{Project} & \textbf{Domain} & \begin{tabular}[b]{@{}c@{}}\textbf{Size}\\\textbf{(KLOC)}\end{tabular} & \textbf{\# Stars} & \begin{tabular}[b]{@{}c@{}}\textbf{\#}\\\textbf{Artifacts}\end{tabular} & \textbf{Project} & \textbf{Domain} & \begin{tabular}[b]{@{}c@{}}\textbf{Size}\\\textbf{(KLOC)}\end{tabular} & \textbf{\# Stars} & \begin{tabular}[b]{@{}c@{}}\textbf{\#}\\\textbf{Artifacts}\end{tabular} \\
    \midrule
    \mbox{mldsa-}\allowbreak\mbox{native} & \mbox{Post-quantum} signatures & 173.6 & 116 & 218 & materialize & Streaming SQL database & 1,089.4 & 6,391 & 6 \\ \addlinespace[1.5pt]
    \mbox{s2n-}\allowbreak\mbox{tls} & TLS and cryptography & 156.5 & 4,783 & 154 & xln & \mbox{Layer-2} payment network & 937.8 & 102 & 28 \\ \addlinespace[1.5pt]
    \mbox{FreeRTOS-}\allowbreak\mbox{Plus-TCP} & Embedded networking & 151.1 & 216 & 99 & \mbox{nonos-}\allowbreak\mbox{microkernel} & Microkernel OS & 517.3 & 300 & 108 \\ \addlinespace[1.5pt]
    \mbox{mlkem-}\allowbreak\mbox{native} & \mbox{Post-quantum} key encapsulation & 116.2 & 254 & 190 & zerocopy & \mbox{Zero-copy} memory conversion & 250.2 & 2,625 & 61 \\ \addlinespace[1.5pt]
    \mbox{FreeRTOS-}\allowbreak\mbox{Kernel} & \mbox{Real-time} OS kernel & 115.1 & 7,933 & 51 & \mbox{s2n-}\allowbreak\mbox{quic} & QUIC transport protocol & 233.6 & 1,377 & 114 \\ \addlinespace[1.5pt]
    \mbox{aws-c-}\allowbreak\mbox{common} & C runtime library & 56.7 & 303 & 173 & \mbox{templar-}\allowbreak\mbox{contracts} & DeFi lending contracts & 225.5 & 11 & 46 \\ \addlinespace[1.5pt]
    \mbox{aws-encryption-}\allowbreak\mbox{sdk-c} & \mbox{Client-side} encryption & 53.4 & 64 & 100 & firecracker & MicroVM monitor & 114.3 & 37,662 & 35 \\ \addlinespace[1.5pt]
    coreMQTT & MQTT messaging client & 33.6 & 234 & 101 & merc & \mbox{Model-checking} toolset & 98.6 & 4 & 23 \\ \addlinespace[1.5pt]
    \mbox{FreeRTOS-}\allowbreak\mbox{Cellular-Interface} & Cellular modem interface & 26.2 & 101 & 63 & \mbox{rust-}\allowbreak\mbox{bitcoin} & Bitcoin protocol library & 73.3 & 2,691 & 5 \\ \addlinespace[1.5pt]
    corePKCS11 & PKCS\#11 cryptographic API & 17.2 & 46 & 30 & spacewasm & WebAssembly interpreter & 71.1 & 1,612 & 72 \\ \addlinespace[1.5pt]
    \mbox{ota-for-aws-}\allowbreak\mbox{iot-embedded-sdk} & IoT \mbox{over-the-air} updates & 14.9 & 66 & 121 & hifitime & Astrodynamics timekeeping & 14.3 & 556 & 153 \\ \addlinespace[1.5pt]
    \mbox{coreMQTT-}\allowbreak\mbox{Agent} & \mbox{Thread-safe} MQTT agent & 5.0 & 46 & 20 & x86\_64 & \mbox{x86-64} hardware abstractions & 9.1 & 942 & 34 \\
    \bottomrule
    \end{tabularx}
\end{table*}

\section{Study Methodology}
\label{sec:methodology}
Figure~\ref{fig:methodology} illustrates our methodology.
We describe our dataset construction and research methods.

\subsection{Dataset Construction}
\label{sec:dataset}

We study BMC artifacts in open-source software projects.
We construct our dataset in four steps.

    \myparagraph{Step 1 --- Selection of BMC Tools}
We focus on two systems-programming ecosystems, C and Rust.
We identified candidate BMC tools through academic publications, tool documentation, and software verification competitions~\cite{bajczi_sv-comp25_2025, beyer_evaluating_2026}.
We shortlisted seven tools and conducted a preliminary GitHub search using tool-specific keywords and code annotations.
Our search identified more repositories for Kani (96) and CBMC (38) than for ESBMC (10), Crux-MIR (8), SeaBMC (4), SMACK (2), and CPAchecker (1).
Hence, we selected CBMC and Kani based on their relatively widespread use and mature industry support.
Although preliminary, these results indicated their adoption in open-source projects.

\myparagraph{Step 2 --- Identification of BMC-based Projects}
We conducted a more extensive search on GitHub and Sourcegraph to identify projects using the selected tools.
For CBMC, our keywords included \verb|__CPROVER_assume|, \verb|cbmc-proof.txt|, and CBMC build configurations.
For Kani, they included \verb|#[kani::proof]|, \verb|#[kani::proof_for_contract]|, and Kani invocation scripts.
These searches yielded 1,602 candidate repositories.

\myparagraph{Step 3 --- Project Filtering and Selection}
We applied two filtering criteria to identify independent software projects containing BMC artifacts.
First, we excluded forks and clones using automated comparisons of repository names, proof filenames, and contents.
Second, we used LLM-assisted screening to exclude tutorials, benchmarks, and non-software repositories.
We manually audited a random sample of exclusions to assess the filtering accuracy.

Of the 457 remaining projects, we selected 12 per tool, prioritizing popularity (GitHub stars), maturity (age and maintainer counts), and application-domain diversity.
\cref{tab:candidate-projects} lists selected projects.
We provide the list of eligible and selected projects and their attributes in the supplementary materials.

\myparagraph{Step 4 --- Identification of Verification Artifacts}
We inspected each selected project to understand how its verification harnesses were organized and executed.
We then developed a script to identify harness entry points, source files, configurations, and associated function contracts.
We treated each distinct combination of these elements as one verification artifact.
For each artifact, we recorded its verification target, source location, and whether it was self-contained or contract-based (\cref{subsec:bg-bmc-harnesses}).
This process identified 2,005 artifacts across 24 projects, summarized in \cref{tab:candidate-projects}.

To assess whether our dataset reflects human-developed BMC practices, we examined artifact commit histories for evidence of AI assistance.
Only three projects (\texttt{zerocopy}, \texttt{spacewasm}, and \texttt{templar-contracts}) showed such evidence, covering 20--90\% of their respective artifacts but only 5.2\% of the full dataset.
These findings suggest that our dataset largely reflects human practices while capturing emerging AI adoption, although unattributed AI use remain undetected.

\subsection{RQ1: Verification Properties}
\label{sec:rq1-method}

Automated BMC systems must select properties that establish meaningful correctness and security guarantees.
RQ1 investigates the properties practitioners verify to inform property inference and identify potential gaps in existing practices.
We distinguish built-in properties configured through tools from custom properties in harnesses or function contracts.

\subsubsection{Built-in Properties}
We reviewed CBMC and Kani's interfaces to identify their supported built-in properties, default settings, and configuration options.
CBMC (v6.x.x) supports 17 properties and Kani supports 11, with eight enabled by default in each tool.
We grouped them into six categories: memory safety (\eg bounds checking), memory resource management (\eg memory leaks), arithmetic safety (\eg integer overflow), value validity (\eg enum range checking), runtime safety (\eg Rust panics), and unwinding assertions.

We developed an automated script to parse each artifact's execution commands and identify enabled properties, accounting for properties enabled by default.
For each category, we measured the proportion of projects enabling some or all properties in some or all of their harnesses.

\subsubsection{Harness-Specific Properties}
We employed LLM-assisted open coding to identify and categorize custom properties specified in verification harnesses and function contracts.
First, one author manually examined harnesses from two CBMC projects and one Kani project, identifying the first two property categories.
The author then used GPT-6.1-sol through Codex to analyze the remaining harnesses in batches of four projects.
The model assigned unambiguous properties to existing categories and flagged new or ambiguous cases for human review.
The author reviewed these cases and introduced or refined categories as needed, retaining responsibility for the taxonomy and its definitions.
While analyzing the harnesses, the model was also tasked to identify and report property specifications whose expression may differ from their intended meaning.
The author reviewed these cases, categorized them and recorded them as property-specification errors.

To assess coding consistency, we selected one CBMC (ota-for-aws-iot-embedded-sdk) and one Kani (spacewasm) project with the greatest diversity of property categories.
An independent GPT-6.1-sol agent classified properties in their harnesses using the resulting taxonomy.
We measured agreement across 193 harnesses and 12 categories, treating each harness-category pair as a binary decision.
The agents agreed on 2,250 of 2,316 decisions (97.15\%; Cohen's $\kappa=0.823$), almost perfect by Landis and Koch's benchmarks~\cite{landis_measurement_1977}.

\subsection{RQ2: Artifact Complexity}
\label{sec:rq2-method}

Automated BMC systems must construct verification artifacts with appropriate input assumptions, function models, and execution bounds.
RQ2 examines the complexity of these artifacts to inform their automated generation and identify opportunities to reduce verification effort.
Building on prior work that uses proof size and the presence of proof constructs to approximate complexity~\cite{matichuk_empirical_2015}, we analyze complexity along three dimensions: harness and input models, function models, and loop bounds.
We use automated scripts to analyze all 2005 artifacts and extract these measurements.

\subsubsection{Harness and Input Models}
We measure harness size and the number of preconditions specified.
For each artifact, we count the lines of code in its verification entry point, excluding function signatures, blank lines, and comments.
We identify preconditions through CBMC's \texttt{\_\_CPROVER\_assume}, Kani's \texttt{kani::assume} and \texttt{kani::any\_where}, and \texttt{requires} clauses in verified function contracts.
Each invocation or contract clause counts as one precondition, regardless of the number of conditions it contains.

\subsubsection{Function Models}
We measure the prevalence and size of custom function models.
We identify three types of function models: (i) functions defined in harness files that match project function signatures or serve as callbacks, (ii) replacement functions configured through Makefiles, and (iii) verified function contracts used to replace implementations.
We exclude harness helper functions and models defined in shared stub files, whose attribution to individual harnesses is difficult to determine.

We statically analyze harness files and Makefiles to identify function models and the proportion of artifacts using them.
For each model, we count its lines of code and precondition clauses.
We report mean models per artifact, model sizes, and precondition counts across projects in each ecosystem.

\subsubsection{Loop Bounds}
Finally, we examine how practitioners configure loop bounds and contracts.
For each artifact, we record whether it specifies global or loop-specific unwinding bounds, enables loop-contract checking, and its maximum configured unwinding bound.
We also identify loops annotated with invariants using constructs such as CBMC's \texttt{\_\_CPROVER\_loop\_invariant(expr)} and Kani's \texttt{\#[kani::loop\_invariant(expr)]}.
We report the proportion of artifacts enabling loop-contract checking, the proportion of projects containing loop invariants, and the number of annotated loops per project.

\subsection{RQ3: Verification Outcomes}
\label{sec:rq3-method}

Automated BMC systems must also produce verification artifacts that achieve meaningful coverage and reliable results at practical execution costs.
RQ3 examines verification results, execution time, and code coverage to establish empirical baselines and identify limitations for verification automation.

\subsubsection{Verification Results}
We developed an automated pipeline to execute each artifact following its project's CI configuration.
We run artifacts individually in Docker containers with a 15-minute timeout, consistent with the time limit adopted by the annual software verification competition~\cite{beyer_evaluating_2026}.
We also impose a 6\,GB memory limit based on available hardware.
We did not increase these limits further because most artifacts completed well within them.
We record whether verification succeeds, reports a property or unwinding assertion violation, or fails due to resource limits or other errors.

\subsubsection{Verification Time}
We measure verification time as the combined duration of symbolic execution and constraint solving, as reported by each tool.
We exclude compilation, instrumentation, and reporting times because these stages differ between CBMC and Kani, limiting comparability.

\subsubsection{Verification Coverage}
We measure code coverage at both artifact and project levels.
We rerun each artifact in coverage mode, using \texttt{cbmc-viewer}~\cite{noauthor_reference_nodate} to process CBMC results and Kani's coverage instrumentation for Rust projects~\cite{noauthor_coverage_nodate}.
For each artifact, we measure line coverage as the proportion of statically reachable lines covered under its input assumptions, function models, and loop configurations, as reported by the tool.
At the project level, we measure function coverage as the proportion of project functions with at least one line covered by any artifact.
Finally, we examine the relationship between verification time and the number of lines covered by each artifact.

\section{Results}

This section presents our findings and interpretations.

\subsection{RQ1: Verification Properties}
\label{subsec:rq1-results}

RQ1 characterizes built-in property adoption, diverse custom verification objectives, and specification errors.

\subsubsection{Built-in Properties}

\begin{figure}
    \centering
    \includegraphics[width=0.8\linewidth]{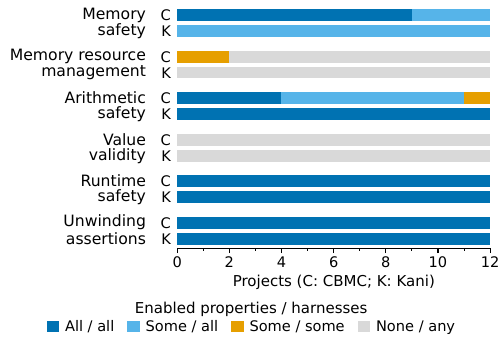}
    \caption{Proportion of CBMC and Kani projects configuring built-in properties. The legend distinguishes projects enabling all or some properties across all or some harnesses.}
    \label{fig:rq1-built-in-properties}
\end{figure}

\Cref{fig:rq1-built-in-properties} shows substantial variation in the adoption of built-in property categories.
Runtime safety checks (CBMC assertions and absence of Rust panics) and unwinding assertions are enabled in all projects.
In contrast, no project enables CBMC's enum range checking or Kani's invalid scalar value checking, while only two CBMC projects enable memory leak checking.
These differences show that available built-in properties are not necessarily adopted.

We also find that default properties are more widely enabled than optional checks.
Across CBMC and Kani projects, 96.9\% and 100\% of default properties are enabled, respectively, compared to 52.8\% and 0\% of non-default properties.
No project explicitly disables a default property, although \texttt{FreeRTOS} lacks some newer default checks because it uses CBMC v5.95.1, predating their introduction in v6.0.0.
These findings suggest that tool defaults and versions strongly influence verified properties, highlighting the importance of explicitly considering built-in checks in automated verification systems.

\begin{finding}{1.1}
Default built-in properties are substantially more widely adopted than optional checks, suggesting that tool defaults strongly influence the properties selected for verification.
\end{finding}

\subsubsection{Custom Harness-Specific Properties}

\input{table/harness-properties}

\Cref{tab:bmc-property-categories} presents our taxonomy of 12 custom property categories and their prevalence across projects.
All Kani projects specify custom properties, compared to 83.3\% of CBMC projects.
Among adopting projects, custom properties appear in an average of 86.5\% of Kani harnesses and 66.1\% of CBMC harnesses.
This widespread use highlights the importance of supporting project-specific specifications beyond built-in safety checks.

Our taxonomy identifies seven property categories in CBMC projects and eleven in Kani projects.
The most prevalent categories concern function outputs: Result Numerical and Pointer Validity (P1), Result Correctness (P2), and Result Structural Validity (P3).
Notably, Result Correctness appears in all Kani projects but only half of CBMC projects, with mean within-project harness shares of 69.2\% and 35.2\%, respectively, among adopting projects.
Furthermore, only Kani projects in our dataset specify State Correctness, Expected Panic Behavior, Function Invocation, Supporting Lemma Correctness, and Model Correctness.
Automated property generators must accommodate these diverse objectives.

\begin{finding}{1.2}
Practitioners specify 12 categories of custom properties beyond built-in safety checks, with substantial differences in verification objectives across the two ecosystems.
\end{finding}

\subsubsection{Case Study on Custom Harness-Specific Properties}

\input{listing/property-categories}

To illustrate these property categories, \cref{lst:rq1-property-examples} presents representative assertions and contracts from the studied projects, organized into four groups.

\myparagraph{Properties of Result Variables (P1--P3)}
These properties constrain values returned by functions or written to output arguments.
For example, a contract in \texttt{mldsa-native} requires polynomial coefficients updated by \texttt{poly\_caddq} to remain within a specified range (P1).
An \texttt{aws-c-common} harness verifies that saturating addition returns the exact sum when representable and the maximum integer otherwise (P2).
Similarly, a \texttt{firecracker} harness checks structural invariants relating a token bucket's budget, capacity, and burst allowance (P3).
These properties range from basic output validity to relationships that establish functional correctness.

\myparagraph{Properties of State and Input Variables (P4--P6)}
These properties establish expectations on program state and the preservation of inputs.
For example, an \texttt{s2n-tls} harness checks that the initialized memory page size falls within a valid range (P4), while a \texttt{spacewasm} harness verifies that allocating a page increases global memory accounting by exactly its size (P5).
A contract in \texttt{mlkem-native} also requires the ciphertext buffer to remain unchanged or be entirely zeroized when encryption fails (P6).
These properties are useful when correctness depends on state changes or preservation rather than function outputs alone.

\myparagraph{Behavioral and Supporting Properties (P7--P10)}
These categories specify exceptional behavior, function interactions, and supporting correctness conditions.
For example, a \texttt{zerocopy} harness verifies that extending an already dynamically sized layout causes a panic (P7), while a \texttt{spacewasm} harness checks that removing and dropping two vector elements invokes their destructors exactly twice (P8).
A \texttt{FreeRTOS-Plus-TCP} harness verifies that ARP processing passes valid buffer pointers to a network-output callback (P9).
Another \texttt{zerocopy} harness proves that multiplying an aligned element size preserves alignment when the product is representable (P10).
These properties capture behavioral requirements and supporting facts beyond output assertions.

\myparagraph{Properties of Verification Artifacts (P11--P12)}
These properties check the validity of verification setups and environment models.
For example, a \texttt{FreeRTOS} harness checks that a configuration bound is positive before initializing queue-lock state (P11).
Similarly, a \texttt{firecracker} harness verifies that its replacement model for \texttt{Instant::now} produces non-decreasing timestamps (P12).
These examples highlight the need to validate verification artifacts themselves, rather than only the software they verify.

\subsubsection{Property Specification Errors}
Our review identified property-specification errors in 35 harnesses across ten projects.
We distinguish two classes based on how they affect verification.
The first involves incorrect property expressions that may alter or weaken the intended guarantees.
These include (E1) aliased snapshots in preservation checks (8 harnesses), (E2) operator-precedence errors (5), (E3) assignments instead of equality checks (3), (E4) comparisons involving incorrect expressions (3), and (E5) incorrect enumeration-membership operators (2).
Other cases involve (E6) checking an unchanged original object instead of its modified clones (1), (E7) an incorrect return-value range predicate (1), and (E8) a tautological Boolean return-value predicate (1).

The second class involves incorrect verification-domain constraints that can exclude relevant executions or make checks vacuous.
These include (E9) incorrect proof-domain predicates that exclude ordinary correctness cases or fail to isolate expected-panic boundary cases (8), and (E10) contradictory or ineffective guards that leave selected assertions unreachable or vacuous (3).
Unlike errors in property expressions, these mistakes can undermine verification even when the assertions themselves are correct.
We provide each confirmed error and its source location in the supplementary materials, counting each affected harness once.
These findings highlight the need for automated techniques that validate both property specifications and the conditions under which they are verified.

\Cref{lst:rq1-property-examples} illustrates three errors.
In \texttt{mlkem-native}, E1 snapshots a pointer rather than its contents, making the preservation branch a self-comparison.
In \texttt{s2n-tls}, E2 parses equality before bitwise OR, so nonzero upper bytes can mask incorrect results.
In \texttt{x86\_64}, E9 is always true for distinct $A,B$, excluding all inputs from the non-panic proof.
All 21 affected harnesses in four projects verify successfully in RQ3.

\begin{finding}{1.3}
Specification errors affect 35 harnesses across ten projects, demonstrating the need to validate both verification properties and their execution conditions.
\end{finding}

\subsection{RQ2: Artifact Complexity}
\label{sec:rq2-results}

Our RQ2 results show that practitioners construct small harnesses with few explicit preconditions, but frequently configure loop bounds and, in some projects, function models.
These characteristics provide empirical baselines for automated harness construction.
\cref{tab:rq2-choices} summarizes our measurements.

\begin{table}[!t]
\centering
\caption{Verification artifact complexity.} %
\label{tab:rq2-choices}
\footnotesize
\input{table/rq2_choices}
\end{table}

\subsubsection{Harness and Input Model Complexity}
Most harness functions are small, with a median of seven lines of code (IQR: 4--14).
Overall, 75\% contain fewer than 15 lines, and only one exceeds 100 lines.
These sizes are comparable to those reported for unit tests~\cite{yu_comprehending_2019,pan_hamster_2026}, while systematically developed unit proofs average 20 lines~\cite{amusuo_unit_2025-1}.
Unlike unit testing, where multiple test cases are often written for each function, practitioners frequently use one harness per verification target.

Input preconditions are also common but generally few.
Although 63\% of artifacts (1,263) specify preconditions, the median is only one per harness, with counts ranging from 0 to 30.
This contrasts with the five preconditions per unit proof reported by Amusuo et al.~\cite{amusuo_unit_2025-1}.
The remaining 742 artifacts (37\%) specify no explicit input preconditions.
This suggests automated harness generation can often begin with relatively simple input models, whose adequacy still requires validation.

\begin{finding}{2.1}
Harnesses are typically small (median: seven lines) and use few explicit input preconditions (median: one), providing practical baselines for automated harness construction.
\end{finding}

\subsubsection{Function Models}
Function modeling varies substantially between the two ecosystems.
Only 5\% of Kani artifacts (34/685) across three projects use custom function models, compared to 44.8\% of CBMC artifacts (591/1,320) across all 12 projects.
Among artifacts using models, the mean is 2.33 per artifact, comparable to the two models per unit proof reported by Amusuo et al.~\cite{amusuo_unit_2025-1}.
For artifacts containing written model bodies, the median total model-body size is nine lines (IQR: 3--14).
These differences may reflect different verification targets or modeling strategies across ecosystems.

Function contracts are the most common form of identified models.
Of the 1,456 models used across artifacts, 725 (49.8\%) replace implementations with verified contracts, while 572 (39.3\%) are defined directly in harness files.
This demonstrates the importance of contract-based abstraction in existing BMC practices and motivates supporting both contract reuse and custom function modeling in automated verification systems.

\begin{finding}{2.2}
Function models are considerably more prevalent in CBMC (44.8\%) than Kani (5\%) artifacts, while verified function contracts account for nearly half of all identified models.
\end{finding}

\subsubsection{Loop Bounds and Contracts}
Explicit loop bounds are common, appearing in 54\% of artifacts (1,082) across 20 projects.
Of these, 62.7\% specify only global bounds, while the others also or exclusively specify loop-specific limits.
The maximum configured bound per artifact has a median of two but reaches 1,515.
These findings highlight the importance of selecting appropriate execution bounds, since insufficient bounds can exclude executions containing property violations.

In contrast, loop contracts are rarely specified.
Although 408 artifacts (20.3\%) enable loop-contract checking, only three projects (\texttt{mlkem-native}, \texttt{mldsa-native}, and \texttt{hifitime}) contain loop invariant annotations, averaging 108.67 clauses per project.
Their limited adoption may reflect the difficulty of specifying suitable invariants~\cite{Furia_2014}.
This contrast motivates automated techniques for inferring loop invariants and reducing reliance on manual unwinding bounds.

\begin{finding}{2.3}
Explicit loop bounds appear in 54\% of artifacts and are typically small (median maximum: two), while loop invariants are specified in only three projects.
\end{finding}

\subsection{RQ3: Verification Outcomes}
\label{sec:rq3-results}

Our RQ3 results show that most BMC artifacts complete verification quickly and cover most of their reachable code, but verification costs vary substantially and repository-level coverage decreases with project size.

\subsubsection{Execution Outcomes}

\begin{table}
    \centering
    \input{table/rq3_outcomes}
    \caption{Verification outcomes per tool: We used a time and memory budget of 900\,s and 6\,GB per run.}    
    \label{tab:rq3-outcomes}
\end{table}

\Cref{tab:rq3-outcomes} shows that 97.2\% (1,948/2,005) of artifacts complete verification within the allocated resource limits.
Of these, 1,946 complete without property violations, while two \texttt{FreeRTOS-Cellular-Interface} artifacts fail only unwinding assertions, indicating insufficient loop bounds.
Of the remaining 57 artifacts, 54 exceed the time or memory limits and one fails to build.

Non-completion is about five times more frequent among Kani artifacts than CBMC artifacts (6.1\% versus 1.1\%), but failures are concentrated in a few projects.
These include \texttt{xln} (12 timeouts among 28 artifacts), \texttt{zerocopy} (12 proofs disabled in CI for being too slow), and \texttt{hifitime} (5 memory-limit failures).
Among the incomplete artifacts, 22 involve potentially expensive execution spaces, including unbounded loops or recursion and symbolically sized buffers.
These findings motivate automated BMC techniques that reduce reliance on finite loop and recursion unwinding through inferred loop invariants, recursive function summaries, and sound abstractions for symbolically sized buffers.

\subsubsection{Runtime and Resource Consumption}

\begin{figure}[!t]
\centering
\includegraphics[width=0.8\columnwidth]{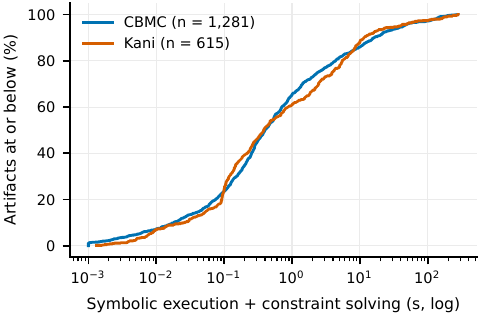}
\caption{Verification-time CDF for completed artifacts.}
\label{fig:rq3-time-ecdf}
\end{figure}

Most completed artifacts require little verification time.
As shown in \cref{fig:rq3-time-ecdf}, the median symbolic execution and constraint-solving time is 0.40\,s for CBMC and 0.38\,s for Kani.
Overall, 64\% complete these phases within one second, 87\% within ten seconds, and 97\% within one minute.
Including overheads, the median wall-clock time is 0.62\,s (CBMC: 0.56\,s; Kani: 0.82\,s).

Memory consumption differs more substantially between the tools.
CBMC has a median peak memory consumption of 80\,MB and a 90th percentile of 0.54\,GB, compared to 376\,MB and 2.5\,GB for Kani.
Kani's higher memory consumption may partly explain its higher non-completion rate, particularly for memory-limited artifacts.
Executing the dataset, including compilation and coverage measurements, required 25.5 machine hours, with another 18.6 hours for the extended runs.

\begin{finding}{3.1}
Most artifacts (97.2\%) complete within the resource limits, with a median wall-clock verification time of 0.62\,s. Resource failures are concentrated in a few projects, highlighting the need for targeted verification configurations.
\end{finding}

\begin{figure}[!t]
\centering
\includegraphics[width=0.8\columnwidth]{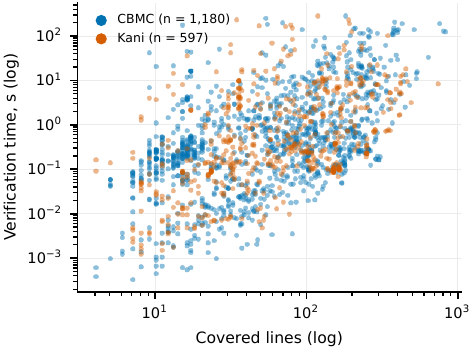}
\caption{Verification time against covered lines per completed artifact, shown on logarithmic axes.}
\label{fig:rq3-time-vs-covered}
\end{figure}

We also examine whether verification time can be predicted from the number of lines covered by an artifact.
\Cref{fig:rq3-time-vs-covered} shows a moderate positive correlation between verification time and covered lines (Spearman's $\rho=0.50$).
We fit power-law models for each tool using least squares in log space on 90\% of the artifacts:

\begin{equation}
t_{\mathrm{CBMC}} = 0.0035\,L^{1.22}
\qquad
t_{\mathrm{Kani}} = 0.0056\,L^{1.09}
\label{eq:rq3-time}
\end{equation}

where $t$ is verification time in seconds and $L$ is the number of covered lines.
The fitted exponents suggest slightly superlinear growth for both tools.
However, artifacts covering similar amounts of code differ in verification time by up to four orders of magnitude.
Covered lines explain only a limited share of runtime variation ($R^2=0.27$), with just 23\% of held-out predictions falling within a factor of two.
These findings suggest that factors such as loops, verification properties, and environment models may help predict verification costs.

\begin{finding}{3.2}
Verification time increases slightly faster than linearly with covered code, but code size alone poorly predicts individual artifact runtime ($R^2=0.27$).
\end{finding}

\subsubsection{Artifact Coverage}

Our results show that most verification artifacts cover a large proportion of their reachable code.
Of the 1,896 artifacts completed within the initial budget, we obtained coverage measurements for 1,781.
The remaining 115 lacked coverage results due to CBMC SMT backend errors (77), coverage timeouts (32), and memory exhaustion (6).

\begin{figure}[!t]
\centering
\includegraphics[
    width=0.8\columnwidth,
    trim=0 0 0 0.6cm,
    clip
]{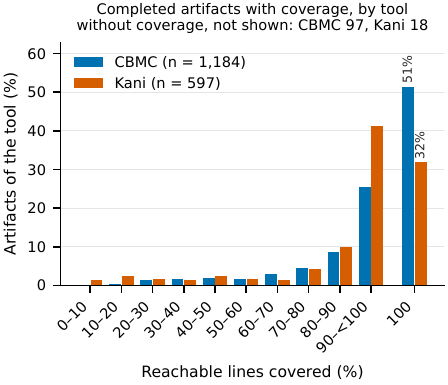}
\caption{Distribution of artifact-level line coverage for CBMC ($n=1{,}184$) and Kani ($n=597$).}
\label{fig:rq3-cov-hist-tool}
\end{figure}

\Cref{fig:rq3-cov-hist-tool} shows a median line coverage of 98\% (CBMC: 100\%; Kani: 96\%).
Overall, 45\% of artifacts cover every reachable line, 24\% cover less than 90\%, and only 7\% (124 artifacts) cover less than half.
Full coverage is more common in CBMC than Kani (51\% versus 32\%), while low coverage is concentrated in \texttt{FreeRTOS-Plus-TCP} (49), \texttt{hifitime} (33), and \texttt{templar-contracts} (15).

We identified four explanations for the 124 artifacts with less than 50\% coverage.
(i) Narrow verification scenarios (63 artifacts, 51\%) use fixed inputs or configurations that leave alternative paths unexplored.
(ii) Coverage measurement artifacts (46, 37\%) include unreachable functions in CBMC's coverage denominator. In 41 \texttt{FreeRTOS-Plus-TCP} artifacts, these are one-line placeholders for excluded functions; removing them raises coverage above 50\% (median 96\%). The remaining five retain functions whose addresses are taken.
(iii) Contract-based abstraction (7, 6\%) leaves callees of replaced function implementations unexecuted in \texttt{hifitime}.
(iv) Vacuous verification (8, 6\%) occurs when execution terminates before reaching the intended assertions.

\subsubsection{Repository Coverage}
\label{sec:rq3-repository-coverage}

Despite high artifact-level coverage, verification artifacts collectively cover a much smaller proportion of larger software projects.
For example, artifacts in \texttt{FreeRTOS-Plus-TCP} statically reach 85\% of its functions but execute at least one line in only 42\%.

We find a strong negative correlation between project size and function coverage (Spearman's $\rho=-0.85$ across 23 projects; CBMC: $-0.76$; Kani: $-0.71$).
Although projects contain between 35 and 12,570 functions, their verification artifacts collectively cover only 1--264 functions per project.
Consequently, smaller libraries often achieve near-complete function coverage, while larger projects cover a substantially smaller fraction of their code.
We exclude \texttt{xln} because its artifacts cover no library functions.

These results reveal that thorough verification within individual artifact scopes does not necessarily translate into broad project-level coverage.
This motivates automated techniques for identifying verification targets and expanding verification breadth, particularly in larger projects.

\begin{finding}{3.3}
Artifacts achieve high coverage within their verification scope (median 98\%), but repository-level function coverage decreases with project size. Low artifact coverage are caused by limited execution scope, coverage tool gaps, and harness errors.
\end{finding}
\section{Discussion}
\label{sec:discussion}

We discuss implications for BMC use and automation.

\subsection{Specifying Meaningful and Correct Properties}
\label{subsec:discussion-properties}

Our RQ1 results show that practitioners verify important software properties through built-in checks and custom specifications.
Built-in checks are widely adopted, particularly when enabled by default, consistent with observations from prior work~\cite{bennett_developers_2024}.
Custom properties also capture consequential requirements: \texttt{mlkem-native} verifies preservation or zeroization of failed cryptographic outputs, \texttt{firecracker} checks rate-limiter invariants, and \texttt{spacewasm} verifies correct memory deallocation.
These examples show that meaningful requirements can be specified compactly and verified efficiently with BMC.

However, uneven adoption raises questions about whether practitioners verify all relevant properties for their projects.
Categories such as State Correctness (P5), Supporting Lemma Correctness (P10), and Model Correctness (P12) appear in few projects, potentially reflecting domain-specific needs, specification difficulty, or limited practitioner awareness.
Prior research on formal verification coverage also highlights that successful verification does not establish the completeness of specified properties~\cite{chockler_coverage_2003}.
Future studies could compare verified properties against requirements and historical defects to identify consequential gaps and develop techniques that recommend useful but overlooked properties.

Beyond selecting useful properties, practitioners must also specify them correctly.
We identified specification errors in 35 harnesses across ten projects, yet 34 still completed verification successfully.
These errors reinforce longstanding concerns about vacuous verification~\cite{beer_efficient_2001} and recent findings that LLM-generated specifications can pass verification without faithfully capturing intended requirements~\cite{wang_are_2026}.
Prior research on vacuity detection and specification coverage~\cite{beer_efficient_2001,chockler_coverage_2003} provides a foundation for techniques that validate property semantics, input feasibility, and assertion reachability.
Independent requirements, historical defects, and mutation testing could further help assess whether properties detect meaningful behavioral violations.

\subsection{Scaling BMC Through Automation}
\label{subsec:discussion-guarantees}

Our RQ2 results show that practitioners construct small harnesses with few preconditions and relatively small loop bounds, suggesting that harness construction may be amenable to automation.
However, it remains unclear whether these choices are sufficient for the intended properties or primarily reflect efforts to reduce verification cost.
Prior studies have shown that incorrect assumptions, specifications, and function models can undermine verification guarantees~\cite{amusuo_unit_2025,priya_verifying_2022,fonseca_empirical_2017}.
How can automated systems make these choices without unnecessarily restricting the behaviors verified?
AutoSOUP~\cite{amusuo2026autosoupsafetyorientedunitproof} provides an initial approach for deriving and validating verification choices for memory-safety proofs, but extending such techniques to the diverse functional properties observed in RQ1 remains an open challenge.

Automation could also address the limited repository-level coverage observed in RQ3, particularly in larger projects.
However, target selection varies: \texttt{FreeRTOS-Kernel} concentrates 65\% of its artifacts on one module, while \texttt{nonos-microkernel} distributes its artifacts across 42 of 52 modules.
Whether these patterns reflect deliberate prioritization or limited verification effort remains unclear, motivating further studies of practitioners' target-selection decisions.
Prior research has explored the feasibility of selecting software analysis targets using  different vulnerability and testing signals~\cite{du_leopard_2019, zhang_intelligen_2021, pecorelli_toward_2024}, but their applicability to verification target selection remains unexplored.
Alternatively, automated harness generation and compositional verification~\cite{beckert_modular_2020} could expand verification breadth, although correctly modeling interactions between components remains challenging.
Future research could evaluate both approaches to determine how best to improve verification guarantees and coverage at practical cost.

\subsection{Threats to Validity}
\label{sec:threats}

\myparagraph{External Validity}
Our 24 selected CBMC and Kani projects may not represent practices across other BMC tools or software projects.
Repository searches and filtering may have missed eligible projects, while differences between ecosystems may reflect programming languages or application domains rather than the tools themselves.

\myparagraph{Construct Validity}
We approximate artifact complexity using code size and construct counts and we exclude shared function models that are difficult to attribute to individual harnesses.
Our LLM-assisted property classification and error identification may also be imperfect despite human review and significant inter-agent agreement.

\myparagraph{Measurement and Interpretation}
Reported high artifact-level coverage does not establish property adequacy or the correctness of verification guarantees. Also, observed associations between artifact characteristics and outcomes do not imply causation.

\section{Conclusion}
We present an empirical study of BMC practices across 2,005 verification artifacts in 24 open-source projects.
Our findings show that practitioners verify diverse correctness and security properties using relatively small artifacts that often complete quickly, but specification errors and limited project-level coverage reveal important limitations.
The property taxonomy, empirical measurements, and annotated dataset we contribute provide a foundation for designing AI-assisted verification systems that specify meaningful properties, construct appropriate verification models, and expand verification breadth.
These advances can enable scalable and trustworthy software assurance as AI increasingly automates software development workflows.

\section{Repository artifact.}
All the scripts, data and results used to produce the results of this paper are published here: \url{https://anonymous.4open.science/r/empirical-study-bounded-model-checking/}

\clearpage
\bibliographystyle{IEEEtran}
\bibliography{bib/main,bib/software-validation,bib/software-verification,bib/software-security}

\end{document}

%% file: typesetting.tex
\usepackage{xspace}
\usepackage{multirow} %
\usepackage{fancyvrb} %
\usepackage{caption}
\usepackage{booktabs} %

\newcommand{\eg}{\textit{e.g.,}\xspace}
\newcommand{\etal}{\textit{et al.}\xspace}

\newif\ifSAVESPACE
\SAVESPACEfalse

\ifSAVESPACE

\else

\fi

\usepackage{todonotes}
\usepackage{soul}

\ifDEBUG
\else
    \newcommand{\AH}[1]{}
    \newcommand{\AM}[1]{}
    \newcommand{\JD}[1]{}
    \newcommand{\TL}[1]{}
    \newcommand{\PA}[1]{}
    \newcommand{\KR}[1]{}
    \newcommand{\LS}[1]{}
    \newcommand{\HP}[1]{}
    \newcommand{\PP}[1]{}
    \newcommand{\RC}[1]{}

\fi

\usepackage{xcolor}
\usepackage{tcolorbox}

\newtcolorbox{finding}[1]{%
  colback=blue!4!white,
  colframe=black!65,
  boxrule=0.6pt,
  arc=2pt,
  left=4pt,
  right=4pt,
  top=3pt,
  bottom=3pt,
  before skip=6pt,
  after skip=6pt,
  fontupper=\small,
  before upper={\textbf{Finding~#1:}\ },
}

\usepackage{hyperref}

\usepackage{cleveref}
\crefformat{section}{\S#2#1#3}
\crefname{figure}{Figure}{Figures}
\crefname{table}{Table}{Tables}
\crefname{theorem}{Theorem}{Theorems}
\crefname{thm}{Theorem}{Theorems}
\crefname{lemma}{Lemma}{Lemmata}
\crefname{equation}{Eqt.}{Eqts.}
\crefformat{Grammar}{Grammar #1}
\crefname{appendix}{Appendix}{Appendices}
\crefname{listing}{Listing}{Listings}
\crefname{algorithm}{Algorithm}{Algorithms}

\newcommand{\myparagraph}[1]{\paragraph{#1}}
\renewcommand{\myparagraph}[1]{\vspace{0.25em} \noindent \hspace{0.15cm}\underline{\textit{#1:}}\xspace}

\usepackage{pifont}

\usepackage{pifont}%

\ifCUSTOMWATERMARK
\usepackage{eso-pic}
\usepackage{graphicx}
\usepackage{xcolor}

\AddToShipoutPictureFG{%
  \AtPageCenter{%
    \put(-0.455\paperwidth,0){%
      \makebox(0,0){%
        \rotatebox{90}{%
          \textcolor{gray!55}{%
            \sffamily\Large
            Draft --- \today
          }%
        }%
      }%
    }%
  }%
}
\fi

%% file: listing/bmc-artifacts.tex
\begin{figure}[t]
  \centering
  \captionsetup{type=lstlisting}
  \label{lst:bg-bmc-artifacts}
  \lstset{
    style=bmcprops,
    basicstyle=\ttfamily\fontsize{5.6}{6.8}\selectfont,
    numbers=left,
    numberstyle=\tiny\color{gray!70},
    numbersep=6pt,
    xleftmargin=12pt,
    firstnumber=1,
    breaklines=true,
    columns=fullflexible,
    aboveskip=2pt,
    belowskip=0pt
  }
  \begin{minipage}[t]{0.49\linewidth}
    \centering\scriptsize\textbf{(a) CBMC (C)}
\begin{lstlisting}
// Function contract
unsigned inc(unsigned x)
  __CPROVER_requires(x < 100)
  __CPROVER_ensures(
    __CPROVER_return_value
      == x + 1)
{ return x + 1; }

// Self-contained harness
void check_inc(void) {
  unsigned x = nondet_uint();
  __CPROVER_assume(x < 100);
  unsigned y = inc(x);
  __CPROVER_assert(
    y == x + 1, "inc");
}

// Contract-checking harness
void contract_harness(void) {
  inc(nondet_uint());
}
// --enforce-contract inc
\end{lstlisting}
  \end{minipage}\hfill
  \begin{minipage}[t]{0.49\linewidth}
    \centering\scriptsize\textbf{(b) Kani (Rust)}
\begin{lstlisting}
// Function contract
#[kani::requires(x < 100)]
#[kani::ensures(
  |r: &u32| *r == x + 1)]
fn inc(x: u32) -> u32 { x + 1 }

// Self-contained harness
#[kani::proof]
fn check_inc() {
  let x: u32 = kani::any();
  kani::assume(x < 100);
  let y = inc(x);
  assert!(y == x + 1);
}

// Contract-checking harness
#[kani::proof_for_contract(
  inc)]
fn contract_harness() {
  inc(kani::any());
}
\end{lstlisting}
  \end{minipage}
\caption{CBMC (left) and Kani (right) examples of self-contained and contract-based verification.}
  
\end{figure}

%% file: table/harness-properties.tex
\begin{table*}[t]
  \centering
  \caption{
  Prevalence of custom BMC property categories. Projects denotes the percentage of adopting projects; Mean H is the mean percentage of harnesses within those projects.
  }
  \label{tab:bmc-property-categories}
  \scriptsize
  \setlength{\tabcolsep}{2pt}
  \renewcommand{\arraystretch}{1.0}
  \setlength{\aboverulesep}{0.25ex}
  \setlength{\belowrulesep}{0.25ex}

  \begin{tabularx}{\textwidth}{
    @{}
    l
    >{\raggedright\arraybackslash}p{0.19\textwidth}
    >{\raggedright\arraybackslash}X
    rrrr
    @{}
  }
    \toprule
    \textbf{ID} & \textbf{Property} & \textbf{What Is Enforced} &
    \multicolumn{2}{c}{\textbf{CBMC}} &
    \multicolumn{2}{c}{\textbf{Kani}} \\
    \cmidrule(lr){4-5}\cmidrule(l){6-7}
    & & & \textbf{Proj.} & \textbf{Mean H} &
    \textbf{Proj.} & \textbf{Mean H} \\
    \midrule

   & \textit{Presence of Custom Properties}
   & \textit{Project-specific requirements in harness assertions or enforced function postconditions.}
   & \textit{83\%} & \textit{66\%} & \textit{100\%} & \textit{86\%} \\

\midrule
\midrule

P1 & Result Numerical or Pointer Validity
   & A result lies within numeric bounds, has an allowed value, or is a valid pointer.
   & 83\% & 32\% & 83\% & 36\% \\

P2 & Result Correctness
   & A result satisfies a relationship with inputs or their pre-execution values.
   & 50\% & 35\% & 100\% & 69\% \\

P3 & Result Structural Validity
   & Result-object fields satisfy relationships required for structural validity.
   & 25\% & 60\% & 67\% & 17\% \\

\midrule

P4 & State Numerical Range
   & A state variable lies within numeric bounds or has an allowed value.
   & 17\% & 2\% & 0\% & \textemdash{} \\

P5 & State Correctness
   & State changes satisfy a specified relationship with function inputs.
   & 0\% & \textemdash{} & 8\% & 6\% \\

P6 & Input and State Preservation
   & Inputs or state remain unchanged or are zeroized, or input objects retain validity.
   & 42\% & 41\% & 42\% & 24\% \\

\midrule

P7 & Expected Panic Behavior
   & An operation triggers an expected panic under specified conditions.
   & 0\% & \textemdash{} & 33\% & 8\% \\

P8 & Function Invocation
   & Another function is invoked, possibly a specified number of times.
   & 0\% & \textemdash{} & 8\% & 3\% \\

P9 & Caller Argument Correctness
   & Arguments passed to another function satisfy required constraints.
   & 42\% & 26\% & 25\% & 2\% \\

P10 & Supporting Lemma Correctness
    & Mathematical facts or implications support an implementation or proof, independently of function results.
    & 0\% & \textemdash{} & 17\% & 20\% \\

\midrule

P11 & Harness and Setup Correctness
    & Checks proof configuration, type sizes or layouts, input construction, or harness-local values.
    & 33\% & 4\% & 58\% & 10\% \\

P12 & Model Correctness
    & Checks function-model behavior or invariants independent of caller-supplied arguments.
    & 0\% & \textemdash{} & 8\% & 3\% \\

    \bottomrule
  \end{tabularx}
\end{table*}

%% file: listing/property-categories.tex
\begin{lstlisting}[
    style=bmcprops,
    float=t,
    caption={Custom properties (P) and specification errors (E).},
    label={lst:rq1-property-examples}
]
// Result variables
(*@\proplabel{P1}@*) ensures(array_bound(a->coeffs, 0,
       MLDSA_N, 0, MLDSA_Q))
(*@\proplabel{P2}@*) if (b > 0 && a > UINT64_MAX - b)
       assert(r == UINT64_MAX);
     else assert(r == a + b);
(*@\proplabel{P3}@*) assert!(bucket.is_valid());

// State and input variables
(*@\proplabel{P4}@*) assert(s2n_mem_get_page_size() > 0);
     assert(s2n_mem_get_page_size()
            <= UINT32_MAX);
(*@\proplabel{P5}@*) assert_eq!(new_total,
       old_total + page_size as isize);
(*@\proplabel{P6}@*) ensures(return_value != 0 ==>
       array_unchanged_or_zeroized_u8(
           c, MLKEM_INDCPA_BYTES))

// Behavioral and supporting properties
(*@\proplabel{P7}@*) #[kani::should_panic]
     let _ = base.extend(field, packed);
// drop_count counts Droppable::drop() calls
(*@\proplabel{P8}@*) assert_eq!(drop_count, 2);
(*@\proplabel{P9}@*) __CPROVER_assert(
       pxNetworkBuffer != NULL, "...");
(*@\proplabel{P10}@*) assert_eq!(bytes & align_mask, 0);

// Verification artifacts
(*@\proplabel{P11}@*) __CPROVER_assert(
        PRV_UNLOCK_QUEUE_BOUND > 0, "...");
(*@\proplabel{P12}@*) assert!(early <= late);

// Specification errors (names abbreviated)
(*@\proplabel{E1}@*) ensures(forall(i, 0, len, r[i] ==
       (b == 0 ? x[i] : old(r)[i])))
(*@\proplabel{E2}@*) assert((hi << 8) | lo == *dest);
(*@\proplabel{E9}@*) let p = addr != A || addr != B; // A != B
     kani::assume(p == panic_mode);
\end{lstlisting}

%% file: table/rq2_choices.tex
\begingroup
\setlength{\tabcolsep}{4pt}
\begin{tabularx}{\columnwidth}{@{}>{\raggedright\arraybackslash}Xrr@{}}
\toprule
\textbf{Metric} & \textbf{CBMC} & \textbf{Kani} \\
\midrule
\multicolumn{3}{@{}l}{\emph{Harness and input models}} \\
\quad Mean harness size (LOC) & 10.34 & 11.12 \\
\quad Projects with input preconditions & 12/12 (100.0\%) & 10/12 (83.3\%) \\
\quad Harnesses with input preconditions & 1,039 (78.7\%) & 224 (32.7\%) \\
\quad Mean preconditions / adopting harness & 3.00 & 1.74 \\
\midrule
\multicolumn{3}{@{}l}{\emph{Function models}} \\
\quad Projects using function models & 12/12 (100.0\%) & 3/12 (25.0\%) \\
\quad Harnesses using function models & 591 (44.8\%) & 34 (5.0\%) \\
\quad Mean models / adopting harness & 2.37 & 1.71 \\
\quad Contract replacements (\% of models) & 48.8\% & 74.1\% \\
\midrule
\multicolumn{3}{@{}l}{\emph{Loop bounds and contracts}} \\
\quad Projects with explicit numeric bounds & 10/12 (83.3\%) & 10/12 (83.3\%) \\
\quad Harnesses with explicit numeric bounds & 909 (68.9\%) & 173 (25.3\%) \\
\quad Mean largest bound / adopting harness & 15.21 & 21.50 \\
\quad Projects with explicit loop invariants & 2/12 (16.7\%) & 1/12 (8.3\%) \\
\bottomrule
\end{tabularx}
\endgroup

%% file: table/rq3_outcomes.tex
\begin{tabular}{lrr}
\toprule
Metric & CBMC & Kani \\
\midrule
\multicolumn{3}{l}{\emph{Completion}} \\
\quad Completed verification & 1,305 (98.9\%) & 643 (93.9\%) \\
\quad\quad within the main budget & 1,281 (97.0\%) & 615 (89.8\%) \\
\quad\quad only with the extended budget & 24 (1.8\%) & 28 (4.1\%) \\
\midrule
\emph{Did not complete} & 15 (1.1\%) & 42 (6.1\%) \\
\quad Timeout (verification budget) & 1 (6.7\%) & 24 (57.1\%) \\
\quad Out of memory & 12 (80.0\%) & 17 (40.5\%) \\
\quad Build error & 1 (6.7\%) & 0 (0.0\%) \\
\quad Required solver not installed & 1 (6.7\%) & 0 (0.0\%) \\
\quad Unsupported language feature & 0 (0.0\%) & 1 (2.4\%) \\
\midrule
\multicolumn{3}{l}{\emph{Unsuccessful verdict}} \\
\quad Property violation & 0 (0.0\%) & 0 (0.0\%) \\
\quad Only unwinding-assertion failures & 2 (0.2\%) & 0 (0.0\%) \\
\bottomrule
\end{tabular}